\documentclass[a4paper,11pt,nohyper]{JHEP3}
\usepackage{amsmath,latexsym,amssymb}
\usepackage{graphicx}
\usepackage[latin1]{inputenc}
\usepackage{subfigure}
\usepackage{nicefrac}

\usepackage{comment}

\newcommand{\slsh}[1]{\displaystyle{\not} #1}
\newcommand\nn{\nonumber}
\newcommand\g{\gamma}
\newcommand\beal{\begin{align}}
\newcommand\tr{\mathrm{tr}}
\newcommand{\boxedeq}[1]{
\begin{equation}
\fbox{
\rule[0.7cm]{0pt}{0pt}
$#1$
\rule[-0.45cm]{0pt}{0pt}
}
\end{equation}
}

\newcommand{\spl}[1]{\begin{split}#1\end{split}}

\title{A note on the quartic effective action\\ of type IIB superstring}

\author{Giuseppe Policastro${}^{\diamondsuit}$ and Dimitrios Tsimpis${}^{\clubsuit}$ \\

  \begin{itemize}
  
\item Laboratoire de Physique Théorique de l'Ecole Normale Supérieure\footnote{Unité
      mixte du CNRS et de l'Ecole Normale Supérieure, UMR 8549.}\\
      24 rue Lhomond, 75231 Paris Cedex 05, France
  
\item  Arnold-Sommerfeld-Center for Theoretical Physics\\
Department für Physik, Ludwig-Maximilians-Universität München\\
Teresienstra\ss e 37, 80333 München, Germany
  \end{itemize}

\bigskip
 E-mail:
\email{policast@lpt.ens.fr, dimitrios.tsimpis@lmu.de}}

\abstract{\textsc We recast the result of our recent computation of the quartic 
action of ten-dimensional  IIB supergravity  in a manifestly SL(2,R)-covariant 
form. This affords us  a critical assessment of an earlier conjecture in the literature. }

\preprint{ LMU-ASC 02/09\\LPTENS-08/69\\}

\keywords{IIB string theory, higher-order corrections}

\begin{document}

\setcounter{footnote}{0}
\renewcommand{\thefootnote}{\arabic{footnote}}
\setcounter{section}{0}


\section{Introduction}\label{sec:intr}

In a recent paper \cite{pt} 
we computed the full four-point amplitude of both type II  
ten-dimensional superstrings to all orders in $\alpha'$, at tree level in the string coupling $g_{\mathrm{s}}$. 
Higher-order derivative corrections to ordinary type IIB supergravity continue to attract a lot of interest. 
Recent applications range from  branes and black holes \cite{hana,hung,haro},  to 
holographic hydrodynamics and QCD \cite{bucha,buchb,basu,yaffe,paul}, to the question of 
perturbative finiteness of $d=4$, $\mathcal{N}=8$ supergravity  \cite{gree}.

The result of \cite{pt}, reviewed below in section \ref{sec:lagr}, was compactly expressed in 
terms of traces of gamma matrices. However, although there was a brief discussion 
concerning the quadratic axion-dilaton terms,  in the form given in \cite{pt}
 the $SL(2,\mathbb{R})$ covariance of the IIB action was not manifest.

On the other hand  a $g_{\mathrm{s}}$-exact conjecture 
for the quartic part of ten-dimensional IIB superstring effective action at order $\alpha^{\prime3}$ 
 (eight derivatives) was put forward some time ago in \cite{kh}. 
The conjecture of \cite{kh} was a -- by no means unique --   
$SL(2,\mathbb{Z})$-invariant extension of the NSNS terms known at the time. 
In the subsequent paper \cite{kp2} it was explained how the action of \cite{kh} is generated by 
taking the orbit of the NSNS terms under the action of (a quotient of) $SL(2,\mathbb{Z})$, 
and it was also noted that additional string calculations would be needed to determine the 
tree-level interactions of RR fields.

Although there is overlap with the computation of \cite{pt}, there is still some confusion 
in the literature concerning the status of the conjecture of \cite{kh}, since the form in which the result of \cite{pt} 
was expressed does not lend itself to immediate comparison with the conjecture of \cite{kh}. 
In the present paper we recast the computation of \cite{pt} in a manifestly $SL(2,\mathbb{R})$-covariant form 
which can be readily compared with \cite{kh}.

Our main result is 
contained in eqn.~(\ref{thelagr}) below. 
As in \cite{kh}, we have not included the terms containing the RR five-form, and we have 
truncated at order $\alpha^{\prime3}$. 
However, we stress that eqn.~(\ref{thelagr}) is only part of the 
quartic action, which can be found in its completeness in \cite{pt}. 
For example, the all-order $\alpha'$-dependence  can be 
easily recovered by acting on the right-hand-side with the operator $\widehat{\mathcal{G}}$,
 as described in detail in \cite{pt}.

In contrast to the $SL(2,\mathbb{Z})$-invariant Lagrangian presented in eqn.~(4.1) of \cite{kh}, eqn.~(\ref{thelagr}) below 
does not contain any higher-$g_{\mathrm{s}}$ corrections. Nevertheless, the two Lagrangians can be 
readily compared at 
tree-level, i.e. in the weak-coupling limit. Indeed, as explained in section \ref{sec:sltz}, 
there are several discrepancies 
between eqn.~(4.1) of \cite{kh} and eqn.~(\ref{thelagr}) of the present paper. 
We therefore conclude that the conjecture of \cite{kh} cannot be the correct $SL(2,\mathbb{Z})$-invariant quartic Lagrangian. It would be interesting to apply the methods of \cite{kp2} in order to determine 
 the full $SL(2,\mathbb{Z})$-invariant extension of the quartic Lagrangian. 

The remainder of this paper is organized as follows: In the next section we review the result of \cite{pt}, which 
we then recast in a manifestly $SL(2,\mathbb{R})$-covariant form in section \ref{sec:sltz}. Appendix \ref{sec:conv} 
contains some useful formul\ae. Many details of our computation are contained in appendix \ref{details}.


\section{Review of the quartic action}\label{sec:lagr}

This section contains a brief review of the result of \cite{pt}, which the reader may consult for further details. 
For ease of comparison with \cite{kh}, in the following we will set $\alpha^\prime=1$ 
and $\kappa^2=1/2$. The complete tree-level 
four-point eight-derivative bosonic 
Lagrangian consists of the following pieces \cite{pt}:

$\bullet$ {NS-NS}
\beal
{\cal L}_{{NS-NS}}= \frac{\zeta(3)}{3\cdot 2^{8}} 
t_8t_8 \widehat{R}^4
 ~,
\label{pipa}
\end{align}
where 
\beal
\widehat{R}_{mn}{}^{pq}:=R_{mn}{}^{pq}
+\sqrt{2} e^{-\frac{ D}{{2}}}\nabla_{[m}H_{n]}{}^{pq}
-\delta_{[m}{}^{[p}\nabla_{n]}\nabla^{q]}D
~.
\label{modifiedconnection}
\end{align}

$\bullet$ {$(\partial F)^2R^2$}
\beal
{\cal L}_{{(\partial F)^2R^2}}= \sqrt{2} \zeta(3)
 (A_1+\frac{1}{2}A_2+\frac{1}{4}A_3)
 ~,
\label{eq:f2r2}
\end{align}
where we have defined
\beal
A_1&:=\widehat{R}^i{}_{n}{}^j{}_{n'}\widehat{R}_{ipjp'}
<\g^n\partial^p\slsh{F}\g^{(n'}\partial^{p')}\slsh{F}^{Tr}>\nn\\
A_2&:=\widehat{R}_{mn}{}^i{}_{n'}\widehat{R}_{pqip'}\Big(
<\g^{mnp}\partial^q\slsh{F}\g^{(n'}\partial^{p')}\slsh{F}^{Tr}>
+<\g^{mnp}\partial^q\slsh{F}^{Tr}\g^{(n'}\partial^{p')}\slsh{F}>\Big)\nn\\
A_3&:=\widehat{R}_{mnm'n'}\widehat{R}_{pqp'q'}
<\g^{[mnp}\partial^{q]}\slsh{F}\g^{m'n'p'}\partial^{q'}\slsh{F}^{Tr}>~.
\label{gamtr}
\end{align}
In the above, $\slsh{F}$ is the Clifford-algebra element:
\beal
\slsh{F}^{\alpha\bar\alpha}=\sum_p \frac{c_p}{p!}(\gamma_{m_1\dots m_p})^{\alpha\bar\alpha}F^{m_1\dots m_p}; ~~~~~
c_p^2=\frac{(-1)^{p+1}}{16\sqrt{2}}~.
\label{fexp}
\end{align}

$\bullet$ {$(\partial F)^4$}
\beal
{\cal L}_{{(\partial F)^4}}= \frac{\zeta(3)}{18 } 
(B_{1}-5 B_{2}+ B_{3}+4 B_{4}- B_{5})
 ~,
\label{pipc}
\end{align}
where we have defined
\beal
B_{1}&:=<\partial_m\partial_p\slsh{F}\g_q\partial^m\partial^p\slsh{F}^{Tr}\g_n\slsh{F}\g^q\slsh{F}^{Tr}\g^n>\nn\\
B_{2}&:=<\partial_m\partial_p\slsh{F}\g_q\slsh{F}^{Tr}\g_n\partial^m\partial^p\slsh{F}\g^q\slsh{F}^{Tr}\g^n>\nn\\
B_{3}&:=<\partial_m\partial_p\slsh{F}\g_q\slsh{F}^{Tr}\g_n\slsh{F}\g^q\partial^m\partial^p\slsh{F}^{Tr}\g^n>\nn\\
B_{4}&:=<\partial_m\partial_p\slsh{F}\g_q\slsh{F}^{Tr}\g_n><\partial^m\partial^p\slsh{F}\g^q\slsh{F}^{Tr}\g^n>\nn\\
B_{5}&:=<\slsh{F}\g_q\slsh{F}^{Tr}\g_n><\partial^m\partial^p\slsh{F}\g^q\partial_m\partial_p\slsh{F}^{Tr}\g^n> \,.\nn\\
\end{align}
%


\section{The covariant four-point Lagrangian}\label{sec:sltz}

In this section we recast the Lagrangian of \cite{pt} in a manifestly 
$SL(2,\mathbb{R})$-covariant form. Putting 
the results of appendix \ref{details} together we have:\footnote{In (\ref{quartic1}) we have reinstated the 
dilaton exponentials according to each field's conformal weight \cite{gs}.}
\beal
\mathcal{L}_{4\mathrm{pt}}&=\frac{\zeta(3)}{3\cdot 2^8}t_8t_8\Big\{R^4+
6R^2\left[ (\partial\partial D)^2+e^{2D}(\partial\partial \chi)^2
+e^{-D}(\partial H^1)^2+e^{D}(\partial H^2)^2\right]
\nn\\
&+6\left[(\partial\partial D)^2+e^{2D}(\partial\partial \chi)^2\right]
\left[ e^{-D}(\partial H^1)^2+e^{D}(\partial H^2)^2\right] \nn\\
&+12R\partial\partial D\left[ e^{-D}(\partial H^1)^2+e^{D}(\partial H^2)^2\right]\Big\}\nn\\
&+{\zeta(3)}\widehat{\mathcal{O}}_1\Big\{\left[ (\partial\partial D)^2+e^{2D}(\partial \partial\chi)^2 \right]^2 \Big\}
+{\zeta(3)}\widehat{\mathcal{O}}_2\Big\{\left[ e^{-D}(\partial H^1)^2+e^{D}(\partial H^2)^2 \right]^2\Big\}
~,
\label{quartic1}
\end{align}
where we are using the shorthand notation:
\beal
t_8t_8\Big\{ABCD\Big\}:=(t_8)^{a_1\dots a_8}(t_8)^{b_1 \dots b_8}
A_{a_1a_2b_1b_2}B_{a_3a_4b_3b_4}C_{a_5a_6b_5b_6}D_{a_7a_8b_7b_8}~,
\end{align}
for any fourth-rank tensors $A$, $B$, $C$, $D$; $\partial\partial D$ is shorthand 
for the fourth-rank tensor $(\partial\partial D)_{ab}{}^{cd}:=\delta_{[a}{}^{[c}\partial_{b]}\partial^{d]}D$; 
$\partial H$ is shorthand 
for the fourth-rank tensor $(\partial H)_{abcd}:=\partial_{[a}H_{b]cd}$; the operators $\widehat{\mathcal{O}}_{1,2}$ 
are defined in eqs (\ref{b60},\ref{b50}) respectively; to facilitate comparison with \cite{kh},  we have 
set $\partial\chi:=\sqrt{2}F_{(1)}$, $H^1:=\sqrt{2}H$, $H^2:=\sqrt{2}F_{(3)}$. 


The Lagrangian above can be written more 
succinctly by using the $SL(2,\mathbb{R})$-covariant field strengths $P$, $G$:
\boxedeq{\label{thelagr}
\spl{
\mathcal{L}_{4\mathrm{pt}}=\frac{\zeta(3)}{3\cdot 2^8}t_8t_8\Big\{R^4 & +
6R^2\left( |\partial P|^2 +|\partial G|^2\right)
+6|\partial P|^2|\partial G|^2
+6R(\partial P+\partial \bar{P})|\partial G|^2\Big\} \, \\
& ~~~~~~~~~~~~~~~~~~~~~~~~~~+{\zeta(3)}\widehat{\mathcal{O}}_1\Big\{(|\partial P|^2)^2 \Big\}
+{\zeta(3)}\widehat{\mathcal{O}}_2\Big\{(|\partial G|^2)^2 \Big\}
~,
}}
where in the linear approximation we are working we have:
\beal
\partial P&= \partial\partial D+ie^D\partial\partial\chi\nn\\
\partial G&=e^{-D/2}\partial H^1-ie^{D/2}\partial H^2~.
\end{align}
Note that we have rescaled $P$ by a factor of two with respect to \cite{kh}. The 
$SL(2,\mathbb{R})$-covariant field-strengths $P$, $G$ are charged under the local $U(1)$ symmetry of IIB, 
with charges $q=2$ and $q=1$ respectively. The conjugate fields $\bar P, \bar G$ have the opposite charges.

Let us make a few comments about the general structure of (\ref{thelagr}):
\begin{itemize}
\item  The NSNS and RR three-forms enter only through the neutral (i.e. with zero  local $U(1)$ charge) 
combination $|\partial G|^2$. The axion and the dilaton enter only through the neutral combination $|\partial P|^2$, except for the last term in the first line. In fact all the terms in 
(\ref{thelagr}) have zero local $U(1)$ charge, except for the 
last one in the first line, which is a sum of therms of charge 2 and -2. 
\item The terms in the second line cannot be expressed using the $t_8t_8$ tensor alone. Indeed, we have 
explicitly checked that the operators $\widehat{\mathcal{O}}_{1,2}$ do {\it not} reduce to $t_8t_8$ when 
acting on $(|\partial P|^2)^2$,   $(|\partial G|^2)^2$.
\end{itemize}
As already mentioned in the introduction, 
eqn.~(\ref{thelagr}) of the present paper and the conjectured 
$SL(2,\mathbb{Z})$-invariant Lagrangian given in eqn.~(4.1) of \cite{kh} can 
be compared in the weak-coupling limit. 
Taking into account that all modular functions $f_k$ in eqn.~(4.1) of \cite{kh} have the same 
$g_{\mathrm{s}}\rightarrow 0$ limit, it can readily be seen that there are several discrepancies 
between the two Lagrangians: all the terms in the Lagrangian of \cite{kh} have the $t_{8} t_{8}$ tensor structure \footnote{The terms with $\varepsilon_{10} \varepsilon_{10}$ are not visible in the linearized approximation.  }, and they have $U(1)$ charges ranging from -4 to 4. For example, it has 
been pointed out that quadratic axion terms of the form $R^2(\partial\chi)^2$ 
contribute to the sub-leading correction to the Debye mass \cite{yaffe}. Whereas these 
terms are zero (by construction) in the $g_{\mathrm{s}}\rightarrow 0$ limit\footnote{Recall that 
the $g_{\mathrm{s}}\rightarrow 0$ limit is holographically dual to the large-$N_c$ limit.} of the Lagrangian of \cite{kh}, 
they are nonzero in eqn.~(\ref{thelagr}) above.

We conclude that the conjecture of \cite{kh} cannot be the correct $SL(2,\mathbb{Z})$-invariant quartic Lagrangian.



\bigskip

\begin{acknowledgments}

We would like to thank Kasper Peeters 
for his prompt help with {\it Cadabra} \cite{cadabra}, and H. Partouche for discussions. D.T. would like to thank 
\'E{}cole Normale Sup\'e{}rieure and the organizers of the 38th Institut d'\'E{}t\'e{} ``Th\'e{}ories de Jauge, Gravit\'e 
et Th\'e{}ories de Cordes'' for hospitality. The work of G.P. is supported by the EU under the contract  PIEF-GA-2008-221026.  
\end{acknowledgments}


\appendix

\section{Useful formul\ae}\label{sec:conv}

The tensor $t_8$ was first introduced in \cite{schw}. Our normalization is such that:
\beal 
t_{abcdefgh} M_{1}^{ab} M_{2}^{cd} M_{3}^{ef} M_{4}^{gh} = & -2\, 
(\tr M_{1}M_{2} \tr M_{3}M_{4} + \tr M_{1}M_{3} \tr M_{2}M_{4} + \tr M_{1}M_{4} \tr M_{2}M_{3}) \nn\\
& + 8 \, (\tr M_{1}M_{2}M_{3}M_{4} + \tr M_{1}M_{3}M_{2}M_{4} + \tr M_{1}M_{3}M_{4}M_{2}) \, ,
\end{align}
for any four antisymmetric matrices $M_1,\dots M_4$. In particular this implies:
\beal
t_8 t_8 R^4=3\times 2^7&\Big\{
R_{abef}R^{fg}{}_{cd}R^{dahe}R_{gh}{}^{bc}
+\frac{1}{2}R_{abef}R^{fgbc}R_{cdgh}R^{heda}\nn\\
&-\frac{1}{2}R_{abef}R^{efbc}R_{cdgh}R^{dagh}
-\frac{1}{4}R_{abef}R^{ef}{}_{cd}R^{dagh}R_{gh}{}^{bc}\nn\\
&+\frac{1}{16}R_{abef}R^{abgh}R_{ghcd}R^{cdef}+\frac{1}{32}(R_{abcd}R^{abcd})^2
\Big\}~.
\end{align}
Another useful identity is the following:
\beal
t_8 t_8 R^2S^2&=  4 R_{a b c d} R_{a b c d} S_{e f g h} S_{e f g h} + 16 R_{a b c d} R_{a b e f} S_{c d g h} S_{e f g h}\nn\\ &+ 8 R_{a b c d} R_{e f g h} S_{a b c d} S_{e f g h} + 8 R_{a b c d} R_{e f g h} S_{a b e f} S_{c d g h}\nn\\ &- 64 R_{a b c d} R_{a b c e} S_{d f g h} S_{e f g h} - 32 R_{a b c d} R_{a b e f} S_{c e g h} S_{d f g h} \nn\\ &- 64 R_{a b c d} R_{a e f g} S_{b h f g} S_{c d e h} - 64 R_{a b c d} R_{e f g h} S_{a b c e} S_{d f g h}\nn\\ 
&- 64 R_{a b c d} R_{a e f g} S_{b h c d} S_{e h f g} + 128 R_{a b c d} R_{a e c f} S_{b g d h} S_{e g f h}\nn\\ 
&+ 256 R_{a b c d} R_{a e f g} S_{b h c f} S_{d g e h} + 128 R_{a b c d} R_{a e c f} S_{b g f h} S_{d h e g}\nn\\ 
&+ 64 R_{a b c d} R_{e f g h} S_{a e c g} S_{b f d h}~,
\end{align}
for any $R$, $S$ with the symmetries of the Riemann tensor.


\section{The individual terms}\label{details}

In this section  we give further details of the derivation of each term in the 
four-point Lagrangian (\ref{thelagr}).

$\bullet$  {$(\partial F_{(1)})^2R^2$}
\label{sec:f12r2}

Let us first examine the $(\partial F_{(1)})^2R^2$ 
couplings. 
Note that in the linearized approximation the equation of motion for $F_{(1)}$ reads 
$\partial^mF_m=0$. In addition, $F_{(1)}$ must be closed by the Bianchi identities. These two 
conditions are equivalent 
to the statement that $\partial_mF_n$ is a traceless symmetric tensor. In the 
Dynkin notation for $D_5$:
$$
\partial_m F_n\sim (20000)~.
$$
Similarly, at the linearized level, the equation of motion for 
the graviton reads $R_{mn}=0$. In addition, the Riemann tensor obeys the 
Bianchi identities $R_{[mnp]q}=0$. Together with the symmetry of the Riemann tensor 
$R_{mnpq}=R_{pqmn}$, these constraints can be expressed compactly as
$$
R_{mnpq}\sim (02000)~.
$$
It follows that in the case 
at hand there are exactly five inequivalent scalars which can be constructed.
In Dynkin notation:
$$
(\partial F_{(1)})^2R^2\sim  (20000)^{2\otimes_s}\otimes (02000)^{2\otimes_s}
\sim 5\times(00000)\oplus\dots
~.
$$
Explicitly, we can choose a basis $I^{F_{(1)}}_1,\dots I^{F_{(1)}}_5$ of these five scalars as follows
\beal
I^{F_{(1)}}_1&:=\partial_{m}F^n\partial_pF^q R^{imjp}R_{injq} \nn\\
I^{F_{(1)}}_2&:=\partial_{m}F_n\partial^pF^q R^{imjn}R_{ipjq}\nn\\
I^{F_{(1)}}_3&:=\partial_{m}F^n\partial_pF^q R^{mpij}R_{ijnq}\nn\\
I^{F_{(1)}}_4&:=\partial_{m}F_i\partial^iF^n R^{mjkl}R_{njkl}\nn\\
I^{F_{(1)}}_5&:=\partial_{i}F_j\partial^iF^j R^{klmn}R_{klmn}~.
\end{align}
In the linearized approximation around flat space we have in addition: 
$R_{mn}{}^{pq}\sim \partial_{[m}\partial^{[p}h_{n]}{}^{q]}$. Taking this 
into account, it is straightforward to show that in this approximation 
the invariants above are not independent, but obey 
\beal
I^{F_{(1)}}_1-I^{F_{(1)}}_2+\frac{1}{2}I^{F_{(1)}}_3+I^{F_{(1)}}_4-\frac{1}{8}I^{F_{(1)}}_5=0~.
\label{lindep}
\end{align}
As we have argued in \cite{pt}, in the linearized approximation around flat space 
there is a relation
\beal
{R}_{mnm'n'}{R}_{pqp'q'}
<\g^{[mnp}\partial^{q]}&F\g^{m'n'p'}\partial^{q'}F^{Tr}>
=\nn\\
&{R}_{mnm'n'}{R}_{pqp'q'}<F\g^{[m'n'p'}\partial^{q']}\partial^{q}F^{Tr}\g^{mnp}>
~.
\end{align}
This can be explicitly verified in the case at hand: a straightforward computation yields
\beal
{R}_{mnm'n'}{R}_{pqp'q'}
<\g^{[mnp}\partial^{q]}F\g^{m'n'p'}\partial^{q'}F^{Tr}>&=64(I^{F_{(1)}}_1-I^{F_{(1)}}_2
+\frac{1}{2}I^{F_{(1)}}_3+\frac{1}{2}I^{F_{(1)}}_4)\nn\\
{R}_{mnm'n'}{R}_{pqp'q'}
<F\g^{[m'n'p'}\partial^{q']}\partial^{q}F^{Tr}\g^{mnp}>&=\nn\\
-64(I^{F_{(1)}}_1&-I^{F_{(1)}}_2+\frac{1}{2}I^{F_{(1)}}_3+\frac{3}{2}I^{F_{(1)}}_4
-\frac{1}{4}I^{F_{(1)}}_5)~.
\end{align}
The expressions on the right-hand sides of the two equations above can indeed be seen to 
be equal, when (\ref{lindep}) is taken into account.

Let us now proceed to the computation of the $(\partial F_{(1)})^2R^2$ couplings. 
These can be read off of (\ref{eq:f2r2}). Explicitly we have:
\beal
\mathcal{L}_{(\partial F_{(1)})^2R^2}=2\zeta(3)I^{F_{(1)}}_1
~,
\label{rep1}
\end{align}
where we have taken (\ref{lindep}) into account. In addition, in order to compare with 
\cite{kh}, we must set: $\partial\chi=\sqrt{2}F_{(1)}$. The above computation can be facilitated considerably with the help of the 
recently-developped symbolic program Cadabra \cite{cadabra}. Noting that in the 
linearized approximation the Riemann tensor reduces to 
$R_{mn}{}^{pq}=4\partial_{[m}\partial^{[p}h_{n]}{}^{q]}$, we have (in momentum space):
\beal
I^{F_{(1)}}_1=\frac{1}{32}t^2u^2
\tilde{\chi}(k_1)\tilde{\chi}(k_2)\tilde{h}_{mn}(k_3)\tilde{h}^{mn}(k_4)+\dots+ \mathrm{permutations}
~,\label{toto}
\end{align}
where $s,t,u$ are the usual Mandelstam invariants, and $\tilde{\chi}$, 
$\tilde{h}_{mn}$ are the momentum transforms of $\chi$, $h_{mn}$, respectively. 
Here and in the following the ellipses denote terms with fewer than four contractions between the 
momenta, i.e. terms with at least one contraction between a momentum and a polarization. Since 
the expressions of these terms are exceedingly lengthy and not very illuminating, we will not present them explicitly.


$\bullet$  {$(\partial^2 D)^2R^2$}\label{sec:d2r2}

The couplings  $(\partial^2 D)^2R^2$ 
come from $t_8t_8\widehat{R}^4$. Similarly to 
the previous case, there are exactly five inequivalent scalars which can be constructed. 
We can choose a basis $I^D_1,\dots I^D_5$ of these five scalars as follows
\beal
I^D_1&:=\partial_{m}\partial^nD\partial_p\partial^qD R^{imjp}R_{injq} \nn\\
I^D_2&:=\partial_{m}\partial_nD\partial^p\partial^qD R^{imjn}R_{ipjq}\nn\\
I^D_3&:=\partial_{m}\partial^nD\partial_p\partial^qD R^{mpij}R_{ijnq}\nn\\
I^D_4&:=\partial_{m}\partial_iD\partial^i\partial^nD R^{mjkl}R_{njkl}\nn\\
I^D_5&:=\partial_{i}\partial_jD\partial^i\partial^jD R^{klmn}R_{klmn}~.
\end{align}
Using (\ref{pipa}),(\ref{modifiedconnection}), we explicitly compute:
\beal
\mathcal{L}_{(\partial^2 D)^2R^2}=\zeta(3)I^D_1
~.
\label{rep2}
\end{align}
Passing 
to the linearized approximation in momentum space, we find
\beal
I^{D}_1=\frac{1}{16}t^2u^2
\tilde{D}(k_1)\tilde{D}(k_2)\tilde{h}_{mn}(k_3)\tilde{h}^{mn}(k_4)+\dots+\mathrm{permutations}
~,
\end{align}
where  $\tilde{D}$ is the momentum transform of $D$.


$\bullet$  {$(\partial H)^2R^2$}\label{sec:h2r2}

The couplings  $(\partial H)^2R^2$ come from $t_8t_8\widehat{R}^4$. 
Using (\ref{pipa}),(\ref{modifiedconnection}), we compute:
\beal
\mathcal{L}_{(\partial H)^2R^2}=
\frac{\zeta(3)}{16}\Big\{&
t^2 u^2\tilde{B}_{d_1 d_2}(k_1) \tilde{B}_{d_1 d_2}(k_2) \tilde{h}_{d_3 d_4}(k_3) \tilde{h}_{d_3 d_4}(k_4)\nn\\  
- 2& t^2 u^2\tilde{B}_{d_1 d_2}(k_1)  \tilde{B}_{d_1 d_3}(k_2) \tilde{h}_{d_2 d_4}(k_3) \tilde{h}_{d_3 d_4}(k_4) \nn\\ 
- 2&t^2 u^2\tilde{B}_{d_1 d_2}(k_1) \tilde{B}_{d_1 d_3}(k_2)  \tilde{h}_{d_3 d_4}(k_3) \tilde{h}_{d_2 d_4}(k_4)\nn\\ 
- 2& u^3 t~\tilde{B}_{d_1 d_2}(k_1) \tilde{B}_{d_1 d_3}(k_2) \tilde{h}_{d_3 d_4}(k_3) \tilde{h}_{d_2 d_4}(k_4) \nn\\ 
-4&t^2 u^2\tilde{B}_{d_1 d_2}(k_1) \tilde{B}_{d_3 d_4}(k_2)  \tilde{h}_{d_1 d_3}(k_3) \tilde{h}_{d_2 d_4}(k_4)\nn\\ 
- 2&t^3 u~\tilde{B}_{d_1 d_2}(k_1) \tilde{B}_{d_3 d_4}(k_2) \tilde{h}_{d_1 d_3}(k_3) \tilde{h}_{d_2 d_4}(k_4) \nn\\ 
-2& t^3 u~\tilde{B}_{d_1 d_2}(k_1) \tilde{B}_{d_1 d_3}(k_2)  \tilde{h}_{d_2 d_4}(k_3) \tilde{h}_{d_3 d_4}(k_4)\nn\\  
- 2&u^3 t~\tilde{B}_{d_1 d_2}(k_1) \tilde{B}_{d_3 d_4}(k_2) \tilde{h}_{d_1 d_3}(k_3) \tilde{h}_{d_2 d_4} (k_4)
\Big\}\nn\\
+&\dots+\mathrm{permutations}~.
\label{rep3}
\end{align}
In the above we have used 
the linearized approximation, $H_{abc}= 3\partial_{[a}B_{bc]}$, for the NSNS three-form, and 
$\tilde{B}$ is the momentum transform of the two-form potential $B$. 
The Riemann tensor is expanded as in (\ref{toto}). To make contact with the notation of 
\cite{kh}, we should substitute: $H_{abc}\rightarrow 1/\sqrt{2} H^1_{abc}$. The computation 
was greatly facilitated by using \cite{gran} to compute the gamma-traces in (\ref{gamtr}), and 
\cite{cadabra} to manipulate the resulting expression in the linearized approximation.


$\bullet$  {$(\partial F_{(3)})^2R^2$}\label{sec:f32r2}

The couplings  $(\partial F_{(3)})^2R^2$  can be computed from (\ref{eq:f2r2}). 
Explicitly we find: 
\beal
\mathcal{L}_{(\partial F_{(3)})^2R^2}=
\frac{\zeta(3)}{16}\Big\{&
t^2 u^2\tilde{C}_{d_1 d_2}(k_1) \tilde{C}_{d_1 d_2}(k_2) \tilde{h}_{d_3 d_4}(k_3) \tilde{h}_{d_3 d_4}(k_4)\nn\\  
- 2& t^2 u^2\tilde{C}_{d_1 d_2}(k_1)  \tilde{C}_{d_1 d_3}(k_2) \tilde{h}_{d_2 d_4}(k_3) \tilde{h}_{d_3 d_4}(k_4) \nn\\ 
- 2&t^2 u^2\tilde{C}_{d_1 d_2}(k_1) \tilde{C}_{d_1 d_3}(k_2)  \tilde{h}_{d_3 d_4}(k_3) \tilde{h}_{d_2 d_4}(k_4)\nn\\ 
- 2& u^3 t~\tilde{C}_{d_1 d_2}(k_1) \tilde{C}_{d_1 d_3}(k_2) \tilde{h}_{d_3 d_4}(k_3) \tilde{h}_{d_2 d_4}(k_4) \nn\\ 
-4&t^2 u^2\tilde{C}_{d_1 d_2}(k_1) \tilde{C}_{d_3 d_4}(k_2)  \tilde{h}_{d_1 d_3}(k_3) \tilde{h}_{d_2 d_4}(k_4)\nn\\ 
- 2&t^3 u~\tilde{C}_{d_1 d_2}(k_1) \tilde{C}_{d_3 d_4}(k_2) \tilde{h}_{d_1 d_3}(k_3) \tilde{h}_{d_2 d_4}(k_4) \nn\\ 
-2& t^3 u~\tilde{C}_{d_1 d_2}(k_1) \tilde{C}_{d_1 d_3}(k_2)  \tilde{h}_{d_2 d_4}(k_3) \tilde{h}_{d_3 d_4}(k_4)\nn\\  
- 2&u^3 t~\tilde{C}_{d_1 d_2}(k_1) \tilde{C}_{d_3 d_4}(k_2) \tilde{h}_{d_1 d_3}(k_3) \tilde{h}_{d_2 d_4} (k_4)
\Big\}\nn\\
+&\dots+\mathrm{permutations}~.
\label{rep4}
\end{align}
As before we have used 
the linearized approximation, $F_{abc}= 3\partial_{[a}C_{bc]}$, for the RR three-form, 
 and $\tilde{C}$ denotes the momentum transform of the two-form potential $C$. 
To make contact with the notation of 
\cite{kh}, we should substitute: $F_{abc}\rightarrow 1/\sqrt{2} H^2_{abc}$.


$\bullet$  {$\partial^2DR^3$}\label{sec:f13r}

The couplings  $\partial^2DR^3$ come from $t_8t_8\widehat{R}^4$. 
Explicitly we find: 
\beal
\mathcal{L}_{\partial^2DR^3}&\propto 
s^2tu\tilde{D}(k_1)  \tilde{h}_{d_1d_2}(k_2) \tilde{h}_{d_2 d_3}(k_3) \tilde{h}_{d_3 d_1}(k_4) 
+\dots+\mathrm{permutations}  \nn\\
&=\frac{stu}{3}(s+t+u)\tilde{D}(k_1)  \tilde{h}_{d_1d_2}(k_2) \tilde{h}_{d_2 d_3}(k_3) \tilde{h}_{d_3 d_1}(k_4)  +\dots+\mathrm{permutations}  \nn\\
&=0~.
\label{rep5}
\end{align}
The last equality above takes into account the kinematic relation $s+t+u=0$.


$\bullet$  {$(\partial^2D)^3R$}\label{sec:f1r3}

The couplings  $(\partial^2D)^3R$ come from $t_8t_8\widehat{R}^4$. 
A direct computation gives: 
\beal
\mathcal{L}_{(\partial^2D)^3R}=0~.
\label{rep6}
\end{align}
%


$\bullet$  {Terms with an odd number of $H$}\label{sec:h1r3}

These couplings are known to vanish \cite{gs,pt}.


$\bullet$  {$(\partial^2 D)^4$}\label{sec:d4}

The couplings  $(\partial^2 D)^4$ come from $t_8t_8\widehat{R}^4$. 
A direct computation gives: 
\beal
\mathcal{L}_{(\partial^2 D)^4}=\frac{\zeta(3)}{64}s^2u^2 
\tilde{D}(k_1)\tilde{D}(k_2)\tilde{D}(k_3)\tilde{D}(k_4)+\dots+\mathrm{permutations} ~.
\label{red4}
\end{align}
In deriving the above we have used the following relations:
\beal
s^4+2s^3u+\mathrm{permutations}=s^4-2s^2u^2+\mathrm{permutations}=0~.
\label{cycls}
\end{align}
These can be straightforwardly shown by cyclicly permuting $s,t,u$, taking 
into account that $s+t+u=0$. For comparison with (\ref{reffku}) below, let us note that 
(\ref{red4}) can also be written as:
\beal
\mathcal{L}_{(\partial^2 D)^4}=\frac{\zeta(3)}{128}(-s^4+t^4+u^4)
\tilde{D}(k_1)\tilde{D}(k_2)\tilde{D}(k_3)\tilde{D}(k_4)+\dots+\mathrm{permutations} ~,
\label{red4prime}
\end{align}
as can be seen using the identity
\beal
s^2u^2+\mathrm{permutations}=\frac{-s^4+t^4+u^4}{2}+\mathrm{permutations}~.
\label{cyclsprime}
\end{align}
%
%


$\bullet$  {$(\partial F_{(1)})^4$}\label{sec:f14}

The couplings  $(\partial F_{(1)})^4$ come from (\ref{pipc}). 
A direct computation gives: 
\beal
\mathcal{L}_{(\partial F_{(1)})^4}=\frac{\zeta(3)}{64}s^2u^2 
\tilde{\chi}(k_1)\tilde{\chi}(k_2)\tilde{\chi}(k_3)\tilde{\chi}(k_4)+\dots+\mathrm{permutations} ~,
\label{reff14}
\end{align}
where as before we have substituted $F_{(1)}\rightarrow\partial\chi/\sqrt{2}$ and we have taken (\ref{cycls}) 
into account. For comparison with (\ref{reffku}) below, note that 
(\ref{reff14}) can also be written as:
\beal
\mathcal{L}_{(\partial F_{(1)})^4}=\frac{\zeta(3)}{128}(-s^4+t^4+u^4)
\tilde{\chi}(k_1)\tilde{\chi}(k_2)\tilde{\chi}(k_3)\tilde{\chi}(k_4)+\dots+\mathrm{permutations} ~,
\label{reff14prime}
\end{align}
as can be seen using (\ref{cyclsprime}).


$\bullet$  {$(\partial F_{(3)})^4$}\label{sec:f34}

The couplings  $(\partial F_{(3)})^4$ come from (\ref{pipc}). 
A direct computation gives: 
\beal
\mathcal{L}_{(\partial F_{(3)})^4}=\frac{\zeta(3)}{64}&\Big\{ 
 t^2u^2\tilde{C}_{ij}(k_1) \tilde{C}^{ij}(k_2) \tilde{C}_{kl}(k_3) \tilde{C}^{kl}(k_4) \nn\\
&+ 2s^2tu\tilde{C}_{ij}(k_1) \tilde{C}_{kl}(k_2) \tilde{C}^{ik}(k_3) \tilde{C}^{jl}(k_4) 
\Big\}
+\dots+\mathrm{permutations} ~.
\label{reff34}
\end{align}
In deriving the above we have made use of the following identities:
\beal
f(s,t,u)&\tilde{C}_{ij}(k_1) \tilde{C}^{ij}(k_2) \tilde{C}_{kl}(k_3) \tilde{C}^{kl}(k_4)+\mathrm{permutations} \nn\\
&=f(t,u,s)\tilde{C}_{ij}(k_1) \tilde{C}^{kl}(k_2) \tilde{C}_{kl}(k_3) \tilde{C}^{ij}(k_4)+\mathrm{permutations}  \nn\\
&=f(u,s,t)\tilde{C}_{ij}(k_1) \tilde{C}^{kl}(k_2) \tilde{C}_{ij}(k_3) \tilde{C}^{kl}(k_4)+\mathrm{permutations} 
~,\label{uida}
\end{align}
for any function $f$ of $s$, $t$, $u$, and
\beal
f(s,t,u)&\tilde{C}_{ij}(k_1) \tilde{C}_{kl}(k_2) \tilde{C}^{ik}(k_3) \tilde{C}^{jl}(k_4)+\mathrm{permutations}\nn\\
&=f(t,u,s)\tilde{C}^{ij}(k_1) \tilde{C}_{ik}(k_2) \tilde{C}_{lj}(k_3) \tilde{C}^{lk}(k_4)+\mathrm{permutations}\nn\\
&=f(u,s,t)\tilde{C}^{ij}(k_1) \tilde{C}_{ik}(k_2) \tilde{C}^{lk}(k_3) \tilde{C}_{lj}(k_4)+\mathrm{permutations}
~.\label{uidb}
\end{align}
For comparison with (\ref{refkra}) below, note that (\ref{reff34}) can also be written as:
\beal
\mathcal{L}_{(\partial F_{(3)})^4}=-\frac{\zeta(3)}{128}&\Big\{ 
2s^2tu \tilde{C}_{d_1 d_2}(k_1)\tilde{C}_{d_3 d_4}(k_2)\tilde{C}_{d_3 d_4}(k_3)\tilde{C}_{d_1 d_2}(k_4)\nn\\
&+tu(t^2+u^2)\tilde{C}_{d_1 d_2}(k_1)\tilde{C}_{d_1 d_2}(k_2)\tilde{C}_{d_3 d_4}(k_3)\tilde{C}_{d_3 d_4}(k_4)\nn\\
&+8stu^2\tilde{C}_{d_1 d_2}(k_1)\tilde{C}_{d_1 d_3}(k_2)\tilde{C}_{d_2 d_4}(k_3)\tilde{C}_{d_3 d_4}(k_4)\Big\}
+\dots+\mathrm{permutations}~.
\label{reff34prime}
\end{align}
%


$\bullet$  {$(\partial H)^4$}\label{sec:f34r}

The couplings  $(\partial H)^4$ come from $t_8t_8\widehat{R}^4$. 
A direct computation gives: 
\beal
\mathcal{L}_{(\partial H)^4}=\frac{\zeta(3)}{64}&\Big\{ 
 t^2u^2\tilde{B}_{ij}(k_1) \tilde{B}^{ij}(k_2) \tilde{B}_{kl}(k_3) \tilde{B}^{kl}(k_4) \nn\\
&+ 2s^2tu\tilde{B}_{ij}(k_1) \tilde{B}_{kl}(k_2) \tilde{B}^{ik}(k_3) \tilde{B}^{jl}(k_4) 
\Big\}
+\dots+\mathrm{permutations} ~.
\label{refh4}
\end{align}
In deriving the above we have taken (\ref{uida}),(\ref{uidb}) into account.
For comparison with (\ref{refkra}) below, note that (\ref{refh4}) can also be written as:
\beal
\mathcal{L}_{(\partial H)^4}=-\frac{\zeta(3)}{128}&\Big\{ 
2s^2tu \tilde{B}_{d_1 d_2}(k_1)\tilde{B}_{d_3 d_4}(k_2)\tilde{B}_{d_3 d_4}(k_3)\tilde{B}_{d_1 d_2}(k_4)\nn\\
&+tu(t^2+u^2)\tilde{B}_{d_1 d_2}(k_1)\tilde{B}_{d_1 d_2}(k_2)\tilde{B}_{d_3 d_4}(k_3)\tilde{B}_{d_3 d_4}(k_4)\nn\\
&+8stu^2\tilde{B}_{d_1 d_2}(k_1)\tilde{B}_{d_1 d_3}(k_2)\tilde{B}_{d_2 d_4}(k_3)\tilde{B}_{d_3 d_4}(k_4)\Big\}
+\dots+\mathrm{permutations}~.
\label{refh4prime}
\end{align}
%


$\bullet$  {$R\partial^2 D(\partial H)^2$}\label{sec:f34ra}

These couplings come from (\ref{pipa}). We find:
\beal
\mathcal{L}_{R\partial^2 D(\partial H)^2}&=-\frac{\zeta(3)}{8}
 s^2tu\tilde{B}_{i}{}^{j}(k_1) \tilde{B}^{ik}(k_2) \tilde{h}_{jk}(k_3) \tilde{D}(k_4) 
+\dots+\mathrm{permutations} ~.
\label{kuku}
\end{align}
%


$\bullet$  {$R\partial^2 D(\partial F_{(3)})^2$}\label{sec:f34rb}

These couplings come from (\ref{eq:f2r2}). We find:
\beal
\mathcal{L}_{R\partial^2 D(\partial F_{(3)})^2}&=-\frac{\zeta(3)}{8}
 s^2tu\tilde{C}_{i}{}^{j}(k_1) \tilde{C}^{ik}(k_2) \tilde{h}_{jk}(k_3) \tilde{D}(k_4) 
+\dots+\mathrm{permutations} ~.
\label{kukub}
\end{align}
%


$\bullet$  {$(\partial^2 D)^2(\partial H)^2$}\label{sec:f34rc}

These couplings come from (\ref{pipa}). We find:
\beal
\mathcal{L}_{(\partial^2 D)^2(\partial H)^2}=-\frac{\zeta(3)}{32}
 tu^3\tilde{B}_{ij}(k_1) \tilde{B}^{ij}(k_2) \tilde{D}(k_3) \tilde{D}(k_4) 
+\dots+\mathrm{permutations} ~.
\label{kukuc}
\end{align}
%


$\bullet$  {$(\partial^2 D)^2(\partial F_{(3)})^2$}\label{sec:f34rd}

These couplings come from (\ref{eq:f2r2}). We find:
\beal
\mathcal{L}_{(\partial^2 D)^2(\partial F_{(3)})^2}=-\frac{\zeta(3)}{32}
 tu^3\tilde{C}_{ij}(k_1) \tilde{C}^{ij}(k_2) \tilde{D}(k_3) \tilde{D}(k_4) 
+\dots+\mathrm{permutations} ~.
\label{kukud}
\end{align}
%
%
%
%
%
%
%
%


$\bullet$  {$(\partial F_{(1)})^2(\partial H)^2$}\label{sec:f34rca}

These couplings come from (\ref{eq:f2r2}). We find:
\beal
\mathcal{L}_{(\partial F_{(1)})^2(\partial H)^2}=-\frac{\zeta(3)}{32}
 tu^3\tilde{B}_{ij}(k_1) \tilde{B}^{ij}(k_2) \tilde{\chi}(k_3) \tilde{\chi}(k_4) 
+\dots+\mathrm{permutations} ~.
\label{kukuca}
\end{align}
%


$\bullet$  {$(\partial F_{(1)})^2(\partial F_{(3)})^2$}\label{sec:f34rdb}

These couplings come from (\ref{pipc}). We find:
\beal
\mathcal{L}_{(\partial F_{(1)})^2(\partial F_{(3)})^2}=-\frac{\zeta(3)}{32}
 tu^3\tilde{C}_{ij}(k_1) \tilde{C}^{ij}(k_2) \tilde{\chi}(k_3) \tilde{\chi}(k_4) 
+\dots+\mathrm{permutations} ~.
\label{kukuda}
\end{align}
%


%
%
%
%
%
%
%
%
%


$\bullet$  {$\partial F_{(1)}(\partial F_{(3)})^3$}\label{sec:f34q}

These couplings come from (\ref{pipc}). 
A direct computation gives: 
\beal
&\mathcal{L}_{\partial F_{(1)}(\partial F_{(3)})^3}\nn\\
&\propto 
\tilde{C}(k_1) \tilde{C}^{ij}(k_2) \tilde{C}_{ik}(k_3) \tilde{C}_{j}{}^k(k_4) 
\Big\{ t^4+u^4+3tu(t^2+u^2)
+ 4t^2u^2
\Big\}
+\dots+\mathrm{permutations}\nn\\
&=0 ~.
\label{reff34q}
\end{align}
The last equality can be seen as follows: interchanging particles 3 and 4 
implies that $t$ and $u$ are also interchanged. Hence the terms in the curly brackets above 
are symmetric under interchanging $3\leftrightarrow 4$. On the other hand, the prefactor 
depending on the polarizations is antisymmetric.


$\bullet$  {$(\partial F_{(1)})^3\partial F_{(3)}$}\label{sec:f34t}

These couplings vanish. This can be seen most easily by a group-theoretical argument. 
Taking the Bianchi identities and lowest-order equations of motion into account, 
$\partial F_{(1)}$, $\partial F_{(3)}$ transform as follows:
$$
\partial F_{(1)}\sim (20000);~~~~~ \partial F_{(3)}\sim(10100)~,
$$
where we are using the Dynkin notation for $D_5$. On the other hand, 
there are no singlets in decomposition of the tensor product 
$$
(20000)^{3\otimes_s}\otimes(10100)
~.
$$


$\bullet$  {$(\partial^2D)^2 (\partial F_{(1)})^2$}\label{sec:i}

These couplings come from (\ref{eq:f2r2}). A direct computation gives:
\beal
\mathcal{L}_{(\partial^2D)^2 (\partial F_{(1)})^2}=\frac{\zeta(3)}{64}(-s^4+t^4+u^4) 
\tilde{D}(k_1)\tilde{D}(k_2)\tilde{\chi}(k_3)\tilde{\chi}(k_4)+\dots+\mathrm{permutations} ~,
\label{reffku}
\end{align}
where we have taken the identity:
\beal
2t^3u+3t^2u^2+2tu^3=\frac{s^4-t^4-u^4}{2}
\end{align}
into account.
%
%
%
On the other hand, a direct computation gives:
\beal
\frac{\zeta(3)}{3\cdot 2^8} &
t_8t_8\big[6(\partial^2D)^2 (\partial F_{(1)})^2 \big]\nn\\
&=\frac{\zeta(3)}{64}(s^4+t^4+u^4) 
\tilde{D}(k_1)\tilde{D}(k_2)\tilde{\chi}(k_3)\tilde{\chi}(k_4)
+\dots +\mathrm{permutations} ~.
\label{reffkur}
\end{align}
Let us define the operator $\widehat{\mathcal{O}}_1$ so that:
\beal
\mathcal{L}_{(\partial^2 D)^2 (\partial F_{(1)})^2}
&=2\zeta(3)\widehat{\mathcal{O}}_1\Big\{(\partial^2 D)^2(\sqrt{2}\partial F_{(1)})^2\Big\}
~,
\label{b60}
\end{align}
where we are using the shorthand 
notation:
\beal
\widehat{\mathcal{O}}\Big\{A B C D\Big\}:=
\widehat{\mathcal{O}}_{e_1a_1\!a_2a_3e_4\!d_1\!d_2d_3}^{e_2b_1b_2b_3e_3c_1c_2c_3}
A^{e_1a_1a_2a_3}B_{e_2b_1b_2b_3}C_{e_3c_1c_2c_3}D^{e_4d_1d_2d_3}
~,
\end{align}
for any fourth-rank tensors $A$, $B$, $C$, $D$. The operator $\widehat{\mathcal{O}}_1$ can be read off 
straightforwardly from (\ref{eq:f2r2}) and 
can be written entirely in terms of products of Kronecker deltas. 
The explicit expression is rather long and not particularly 
illuminating, so we refrain from quoting it here. It can be readily reproduced with the use of a symbolic gamma-matrix program 
such as \cite{gran}.

The important thing to note is that the definition (\ref{b60}) together with 
eqs (\ref{red4prime},\ref{reff14prime},\ref{reffku}) imply the following 
identities:
\beal
\frac{1}{3\cdot 2^8}t_8t_8\Big\{(\partial^2 D)^4\Big\}
&=\widehat{\mathcal{O}}_1\Big\{(\partial^2 D)^4\Big\}\nn\\
\frac{1}{3\cdot 2^8}t_8t_8\Big\{(\partial F_{(1)})^4\Big\}
&=\widehat{\mathcal{O}}_1\Big\{(\partial^2 \chi)^4\Big\}~.
\end{align}
However, we also have:
\beal
\frac{1}{3\cdot 2^8}t_8t_8\Big\{6(\partial^2 D)^2(\partial F_{(1)})^2\Big\}
&\neq2\widehat{\mathcal{O}}_1\Big\{(\partial^2 D)^2(\partial^2 \chi)^2\Big\}
~,
\end{align}
as follows from (\ref{reffku}, \ref{reffkur}). This implies that, as already mentioned in section \ref{sec:sltz}, 
$\widehat{\mathcal{O}}_1$ does {\it not} reduce to $t_8t_8$ when acting on $(|\partial P|^2)^2$.


$\bullet$  {$R \partial^2D (\partial F_{(1)})^2$}\label{sec:}

These couplings vanish, as can be seen by direct computation.


$\bullet$  {$(\partial H)^2 (\partial F_{(3)})^2$}\label{sec:kra}

These couplings come from (\ref{eq:f2r2}). A direct computation gives:
\beal
\mathcal{L}_{(\partial H)^2 (\partial F_{(3)})^2}&=-\frac{\zeta(3)}{64}\Big\{ 
2s^2tu \tilde{B}_{d_1 d_2}(k_1)\tilde{B}_{d_3 d_4}(k_2)\tilde{C}_{d_3 d_4}(k_3)\tilde{C}_{d_1 d_2}(k_4)\nn\\
&+tu(t^2+u^2)\tilde{B}_{d_1 d_2}(k_1)\tilde{B}_{d_1 d_2}(k_2)\tilde{C}_{d_3 d_4}(k_3)\tilde{C}_{d_3 d_4}(k_4)\nn\\
&+8stu^2\tilde{B}_{d_1 d_2}(k_1)\tilde{B}_{d_1 d_3}(k_2)\tilde{C}_{d_2 d_4}(k_3)\tilde{C}_{d_3 d_4}(k_4)\Big\}\nn\\
&+\dots +\mathrm{permutations} ~.
\label{refkra}
\end{align}
On the other hand we compute:
\beal
\frac{\zeta(3)}{3\cdot 2^8}t_8t_8 \left[6(\partial H)^2 (\partial F_{(3)})^2\right]&=
\frac{\zeta(3)}{256}\Big\{ 
4s^2u^2 \tilde{B}_{d_1 d_2}(k_1)\tilde{B}_{d_3 d_4}(k_2)\tilde{C}_{d_3 d_4}(k_3)\tilde{C}_{d_1 d_2}(k_4)\nn\\
&+2t^2u^2 \tilde{B}_{d_1 d_2}(k_1)\tilde{B}_{d_1 d_2}(k_2)\tilde{C}_{d_3 d_4}(k_3)\tilde{C}_{d_3 d_4}(k_4)\nn\\
&+4s^2tu\tilde{B}_{d_1 d_2}(k_1)\tilde{B}_{d_3 d_4}(k_2)\tilde{C}_{d_1 d_3}(k_3)\tilde{C}_{d_2 d_4}(k_4)\nn\\
&+8st^2u\tilde{B}_{d_1 d_2}(k_1)\tilde{B}_{d_1 d_3}(k_2)\tilde{C}_{d_2 d_4}(k_3)\tilde{C}_{d_3 d_4}(k_4)\Big\}\nn\\
&+\dots +\mathrm{permutations} ~.
\label{unbomb}
\end{align}
Analogously to the previous case, we can define the operator $\widehat{\mathcal{O}}_2$ so that:
\beal
\mathcal{L}_{(\partial H)^2 (\partial F_{(3)})^2}
&=2\zeta(3)\widehat{\mathcal{O}}_2\Big\{(\sqrt{2}\partial H)^2(\sqrt{2}\partial F_{(3)})^2\Big\}
~,
\label{b50}
\end{align}
and again we note that the definition (\ref{b50}) together with 
eqs (\ref{reff34prime},\ref{refh4prime},\ref{refkra}) imply the following 
identities:
\beal
\frac{1}{3\cdot 2^8}t_8t_8\Big\{(\partial H)^4\Big\}
&=\widehat{\mathcal{O}}_2\Big\{(\partial H^1)^4\Big\}\nn\\
\frac{1}{3\cdot 2^8}t_8t_8\Big\{(\partial F_{(3)})^4\Big\}
&=\widehat{\mathcal{O}}_2\Big\{(\partial H^2)^4\Big\}~.
\end{align}
However, we also have:
\beal
\frac{1}{3\cdot 2^8}t_8t_8\Big\{6(\partial H)^2(\partial F_{(3)})^2\Big\}
&\neq2\widehat{\mathcal{O}}_2\Big\{(\partial H^1)^2(\partial H^2)^2\Big\}
~,
\end{align}
as follows from (\ref{refkra}, \ref{unbomb}). This implies that, as already mentioned in section \ref{sec:sltz}, 
$\widehat{\mathcal{O}}_2$ does {\it not} reduce to $t_8t_8$ when acting on $(|\partial G|^2)^2$.


$\bullet$  {$\widehat{R}^2 \partial F_{(1)}\partial F_{(3)}$}\label{sec:teleytaion}

A direct computation shows that all these couplings, which come from (\ref{eq:f2r2}), vanish.



\end{document}